\documentclass[12pt]{iopart}
\usepackage{graphicx}
\usepackage{dcolumn}
\usepackage{bm}
\usepackage{hyperref}

\usepackage{cite}

\begin{document}

\title[Dynamics of a vertical water bridge]{Dynamics of a vertical water bridge}

\author{Reza M. Namin}

\address{Department of Mechanical Engineering, Sharif University of Technology, Tehran, Iran.}
\ead{namin@mech.sharif.edu}

\author{Zahra Karimi}

\address{Department of Chemical and Petroleum Engineering, Sharif University of Technology, Tehran, Iran.}
\ead{karimi@che.sharif.edu}

\begin{abstract}
A vertical connection of water is formed when a high voltage electrode is dipped in and pulled out of a container of deionized water. We considered the formation and dynamical characteristics of this vertical water bridge.
For the first time in this field, instabilities were observed in the bridge that led to an oscillatory behaviour which we categorized them into three dynamical regimes. some explanations  were supplied on the physics behind these dynamics. We report the formation of macroscopic droplets during our experiments, which their dynamics revealed that they are electrically charged. In some cases the droplets levitated in the air due to the equality of gravity and electrical force (acting in the opposite direction). Our results shed light on the physics behind this phenomenon and the horizontal water bridge, which explanations regarding its underlying physics have led to controversial theories and discussions before.
\end{abstract}

\pacs{47.65.-d, 47.55.nk}
\submitto{Journal of Physics D: Applied Physics}
\maketitle

\section{Introduction}
Properties of liquids under influence of electric fields have been the subject of many studies mainly due to their applications in different areas. Electrohydrodynamics is the basic physics behind many phenomenon such as Taylor cones and electro sprays
\cite{taylor1964disintegration,fernandez2007fluid,collins2007electrohydrodynamic},
whipping jets
\cite{taylor1969electrically,hohman2001electrospinning,riboux2011whipping},
or non-coalescing drops \cite{bird2009critical} etc.
Electrohydrodynamics enhances fluid control in micro and nano scale applications such as Lab on a Chip \cite{zeng2004principles,diakite2012low,kovarik2009nanofluidics}
and many other Micro Electro Mechanical Systems \cite{ramos2011electrokinetics,tsai2007review,chiarot2011overview,kim2013drop}.
Thus understanding the behaviour of liquids under influence of electric fields is an important task and our approach is to study unexplained Electrohydrodynamic phenomena, an example of which is the formation and properties of the floating water bridge.

The formation of a water connection under the influence of high voltages, commonly known as the floating water bridge was first reported in 1893 by Armstrong \cite{armstrong1893electrical} and has been the interest of many studies since 2007 when the phenomenon was re-discovered by Fuchs et al \cite{fuchs2007floating}.
To understand the underlying physics in the water bridge and describe its stability mechanisms and many other features observed in the experiments, a significant number of recent investigations have been aimed to study the structural changes in the bridge 
\cite{fuchs2009neutron,ponterio2010raman,fuchs2010two,fuchs2009inelastic,fuchs2012investigation,skinner2012structure}. 
Other studies have attempted to explain the features theoretically using classical electrohydrodynamics 
\cite{widom_t,marin_t,saija2010communication,aerov_t,morawetz2012effect,morawetz_t}.
On the other hand, there are theories explaining the phenomenon utilizing quantum electrodynamics
\cite{del2010influence,del2010collective}.
Since Explanations are in a diverse range and still many experiments are unexplained, the governing physics of this phenomena is not completely understood yet. Attempts to quantitatively compare theoretical predictions with experimental results have been very limited due to many complications in the horizontal bridge including its geometrical complexity. For example in our recent experimental investigation \cite{namin2013equilibrium} we showed that contribution of the electrostatic forces and surface tension are likely responsible for the stability of the bridge, however a comprehensive theoretical explanation is still lacking. 
Reviews on this topic have been published which may be referred to for more information \cite{fuchs2010can,woisetschlager2
012horizontal,fuchs2013armstrong}.

In a different geometry, vertical liquid bridging was first discovered by Raco in 1968 \cite{raco1968electrically}; He observed the formation of a liquid column by applying an electric field normal to the interface of specific liquids.
Since then electrohydrodynamic behaviour of current carrying liquids was studied in some cases \cite{hohman2001electrospinning} afterwards the stability of such a systems was a matter of consideration as well, being both theoretically and experimentally studied under reduced gravity conditions \cite{burcham2002electrohydrodynamic}. In 2008 Fuchs et al \cite{fuchs_dynamics} pointed out the formation of vertical floating water bridge. In 2010 Ponterio et al \cite{ponterio2010raman} performed a Raman scattering measurement on vertical and horizontal water bridges, they evaluated that the behaviour of both vertical and horizontal bridges are the same, mentioning the direction of the gravity force as the only difference between them. Understanding the exact behaviour of vertical water bridge will lead us to a better understanding of this class of phenomena and it will be easier due to less geometrical complications compared to a horizontal bridge.

In this letter we report new experiments on the vertical water bridge. We have observed and explained the dynamical behaviour of the bridge when a relatively high resistance was placed in the circuit.  We categorize the oscillatory motion of the bridge in different regimes and also report the formation of electrically charged droplets.

\section{Experimental}
\begin{figure}
 \centering
 \includegraphics[width=80mm]{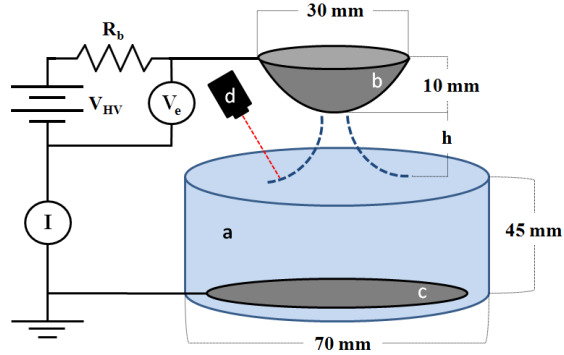}
 \caption{(Color online) Schematic of the experimental setup. (a) beaker filled with deionized water (b) steel spherical electrode (c) aluminium electrode (d) IR thermometer. }\label{fig:setup}
\end{figure}

The experimental setup is shown in Fig. \ref{fig:setup}. A high-voltage power supply was used which was capable of producing a maximum voltage of 25kV and current intensity of 20mA (Plastic Capacitors Inc. HV250-103M). An electrical resistance of $R_b$ was attached to the positive side of the high-voltage power supply as a ballast resistor, which its magnitude was effective in the observed dynamics of the bridge as it will be explained in Results and Discussion. The negative side was connected to the ground and to
an aluminium electrode placed at the bottom of a cylindrical beaker filled with deionized water. The beaker was placed on a jack by which its altitude could be varied. A spherical electrode was placed on top of the beaker,connected to the positive side of the high voltage power supply. Current intensity and electrical voltage drop across the electrodes were measured and monitored using an oscilloscope. An infrared thermometer (TES 1326S) was placed above the beaker in order to measure the surface temperature of water near the bridge, and the temperature was kept between 25 and 35 degrees during the experiments. The deionized water was produced using Millipore Simpack 1 Purification pack kit and initially
had a resistivity of $18.2 M\Omega \cdot cm$. Resistivity of pure DI
water decreases rapidly by contamination of impurities,
e.g. the $CO_2$
gas from air. The resistivity during our
experiments was $1.8 M\Omega \cdot cm$. The resistivity also varies
by temperature changes; it decreases from $1.8 M\Omega \cdot cm$ at
$25^o C$
to $1 M\Omega \cdot cm$ at $45^o C$. A high speed camera capable of recording in 1200 frames per second was used to visualize the bridge from the front view.

\section{Results and Discussion}
\label{sec:res}
\begin{figure*}
 \centering
 \includegraphics[width=30mm]{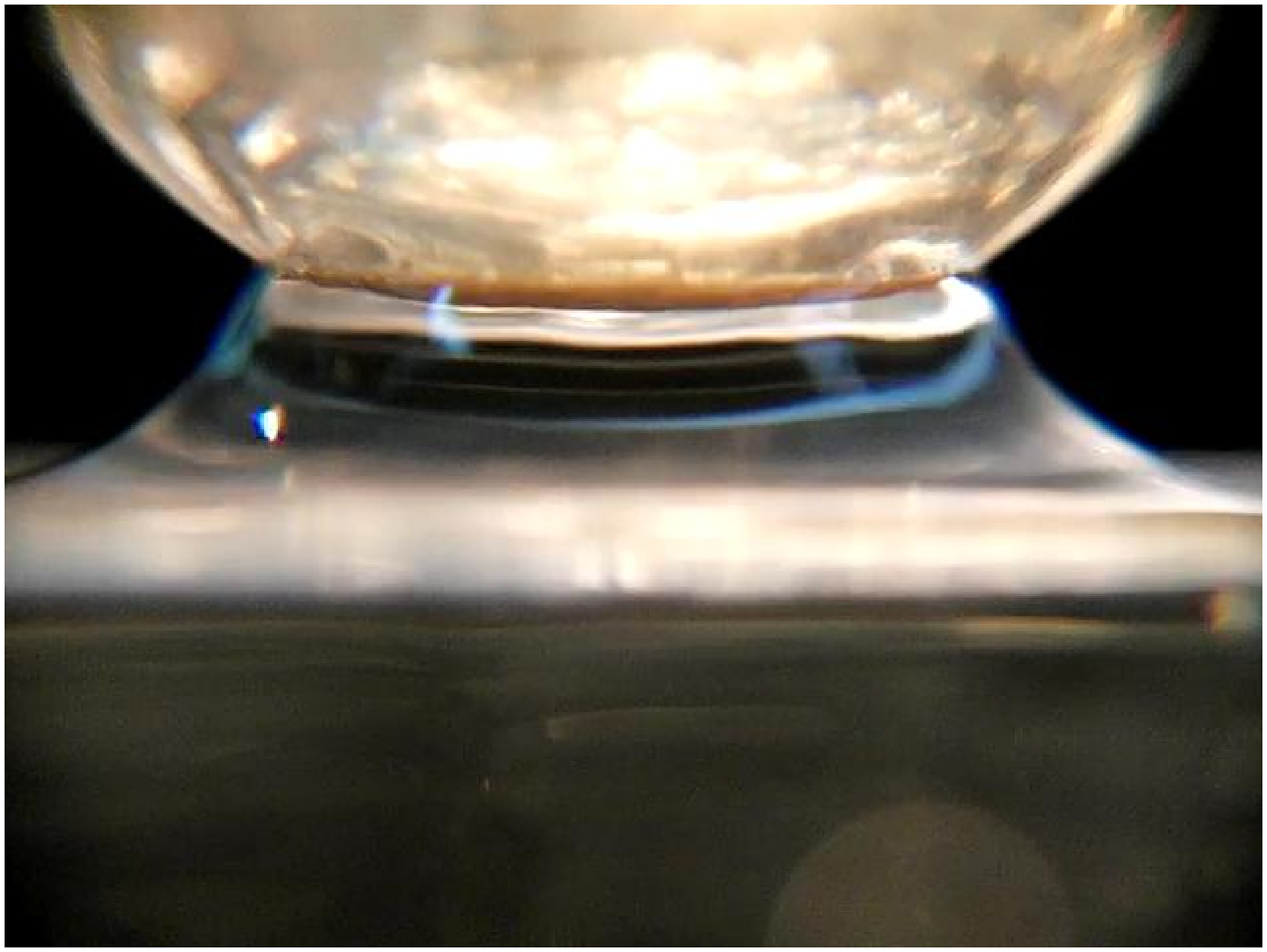}
 \includegraphics[width=30mm]{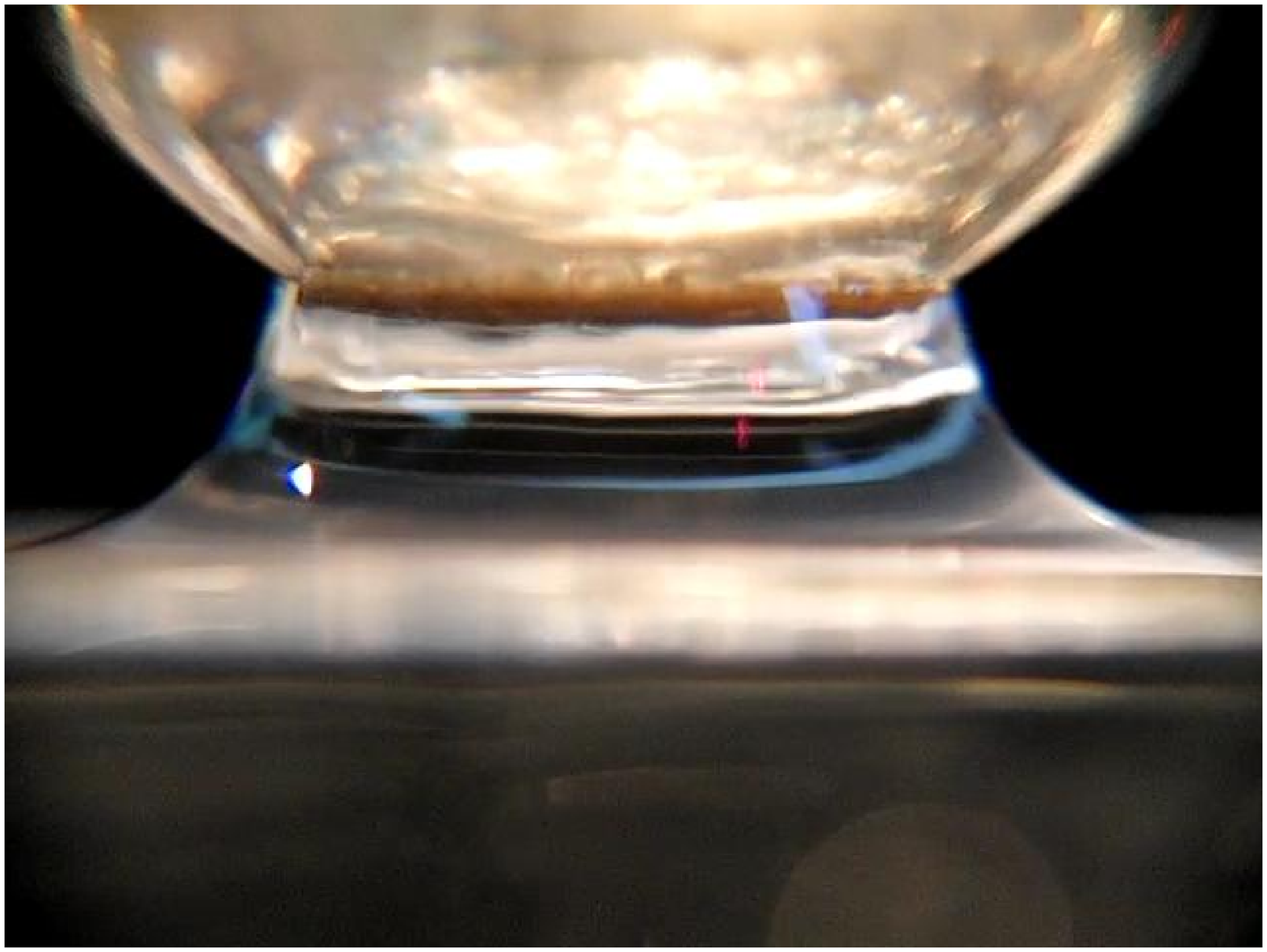}
 \includegraphics[width=30mm]{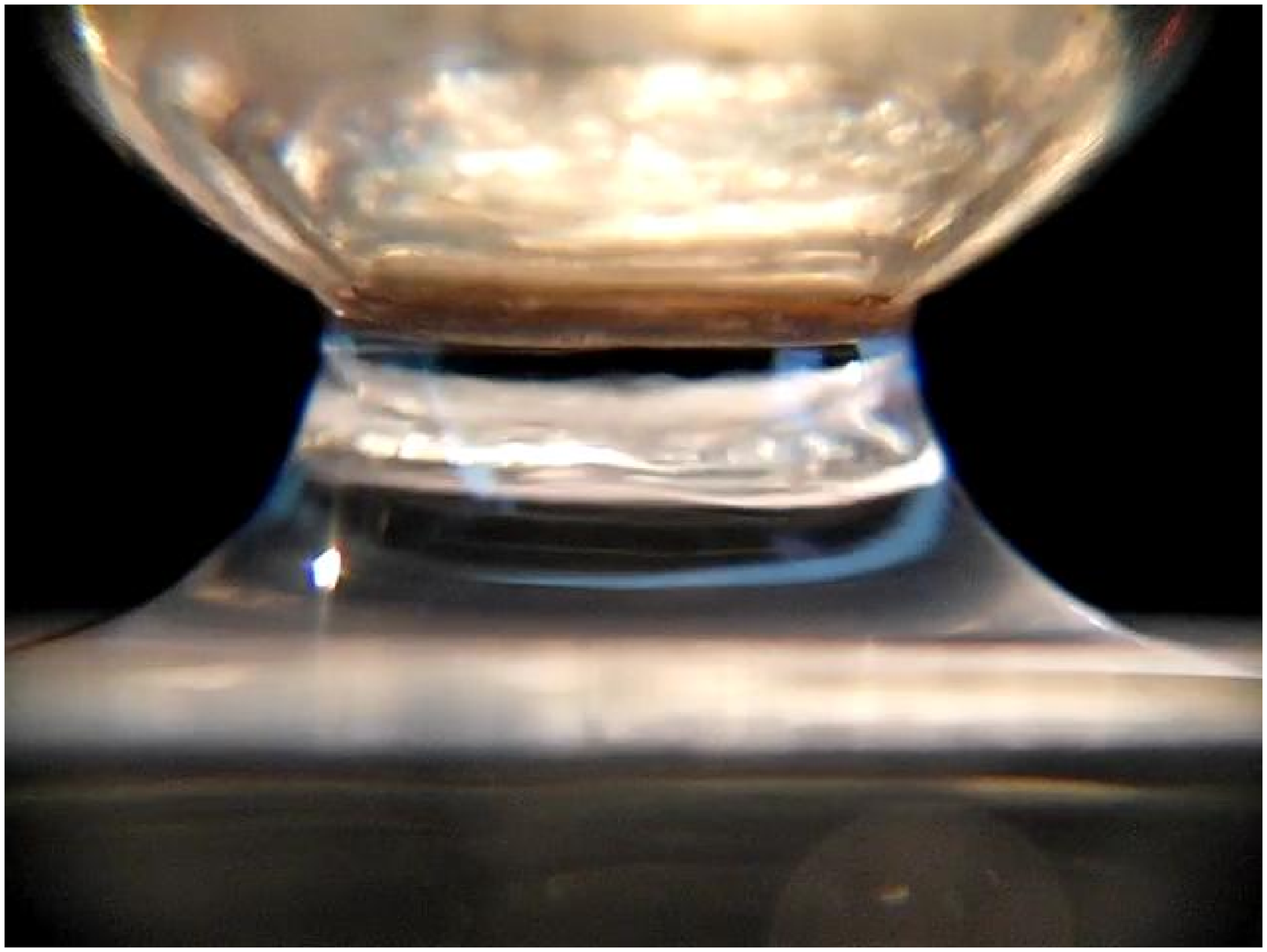}
 \includegraphics[width=30mm]{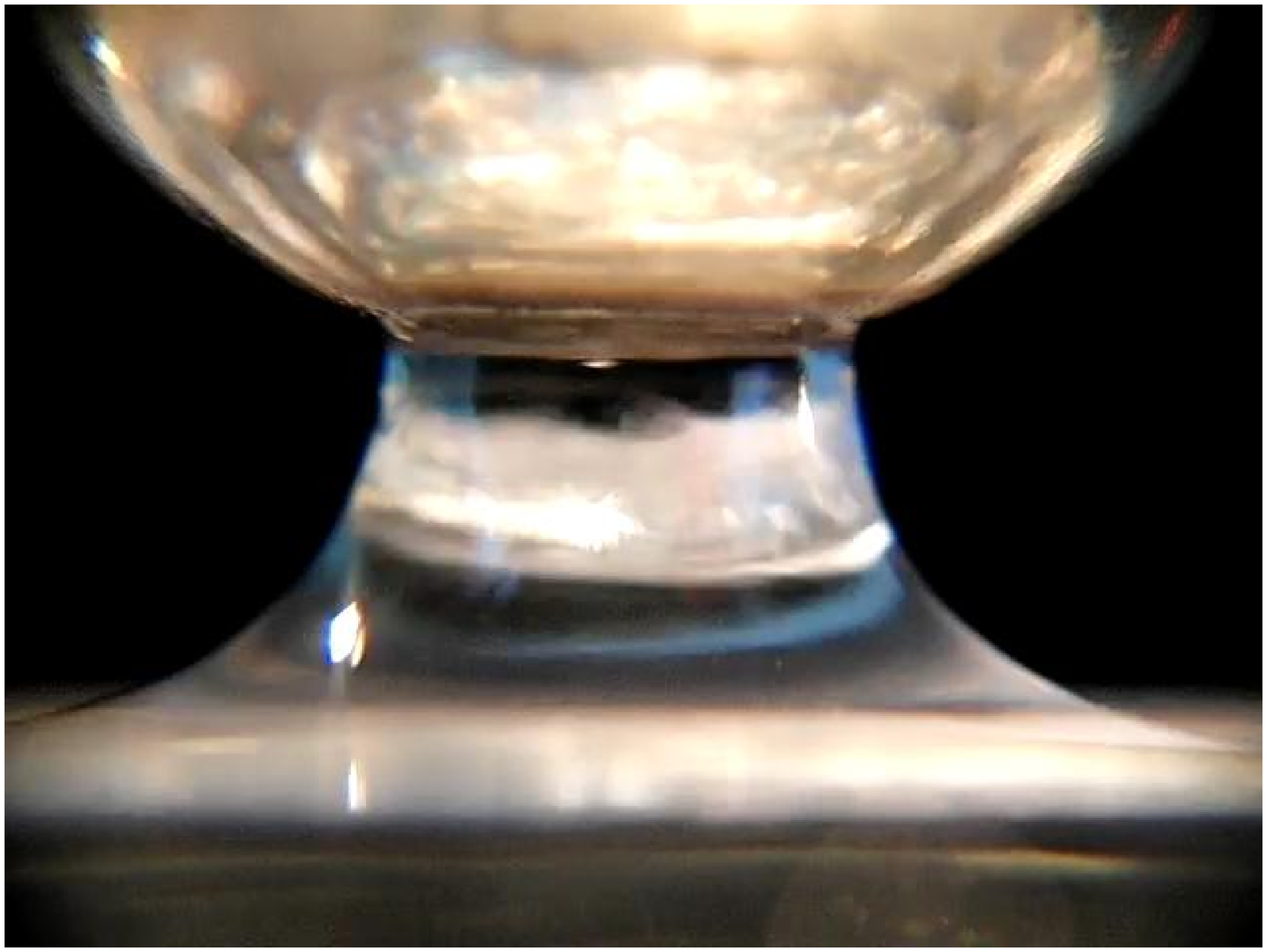}
 \includegraphics[width=30mm]{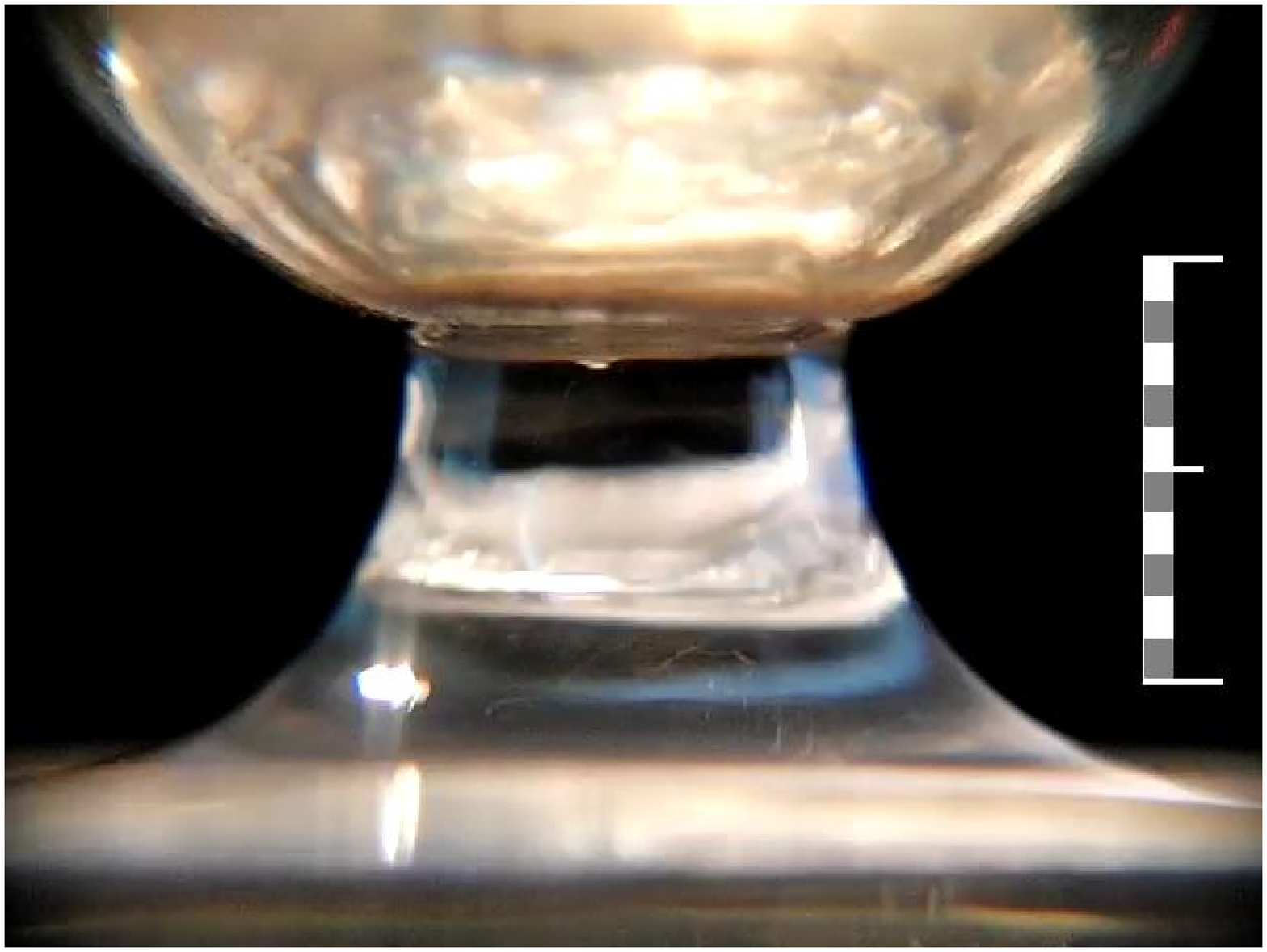}
 \caption{(Color online) The vertical water bridge in a same current intensity of $15mA$ and different heights. Scale of $1cm$ at right.}\label{fig:exp}
\end{figure*}

\begin{figure}
 \centering
 \includegraphics[width=70mm]{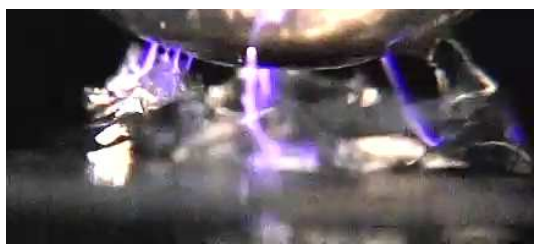}
 \caption{(Color online) Formation of a bridge when a voltage is applied and water is not connected to the electrode. The formation happens after the water surface rises in the form of columns repeatedly and is opposed by electrical discharges. Sometimes the procedure ends to a column finally touching the electrode and diminishing the discharges (Video 1), and sometimes ends to a stable discharge with no water bridge (Video 2).}\label{fig:formation}
\end{figure}

If a relatively low resistance is used as the ballast resistor ($50 k\Omega$), a vertical bridge will be formed by placing the top electrode in contact with water and then pulling it out while the voltage is applied (Fig. \ref{fig:exp}). As the height of the bridge is increased, it gets thinner at its top, then unstable motions occur and finally an electric discharge will be replaced by the bridge. With a voltage of 25kV we could observe vertical bridges as long as 23mm. The formation might also happen when the electrode is near the surface but not in contact with it, so when a voltage is applied, water surface rises up until touching the electrode and then it stabilizes as the bridge (Fig. \ref{fig:formation}). This phenomenon was observed in micro-scale in Atomic Force Microscopy before and is explained by \cite{sacha2006induced}. However in our  case, sparks and discharges occur before the formation of a bridge. The discharges generally reduce the local electric field and avoid the rising jet from touching the electrode. Our high speed visualization shows that it takes a while before the formation of a stable bridge, during which many columns of water rise but discharges are formed and avoid a stable bridge until one column finally reaches the tip and stability is achieved. The procedure is quite chaotic and unreproducible, as for our experiments with similar conditions sometimes a bridge is formed after a while (Video 1) and sometimes a stable discharge takes place (Video 2).

We observed that increasing the resistance of ballast resistor and limiting the current intensity leads to interesting unsteady and unstable bridges. The ballast resistor was $25 M\Omega$ and when the bridge was formed, increasing its height did not lead to a stable bridge. Our observations showed three different regimes of motion in this case: (a) in small bridge heights the bridge varies in its diameter; an oscillatory thinning-thickening motion with a well defined frequency is exhibited (Video 3). Also the bridge frequently moved horizontally and this motion increased its instability, sometimes leading to destruction of the bridge. However it was reconstructed immediately by another rising column of water and the motion continued.

\begin{figure}
 \centering
 \includegraphics[width=40mm]{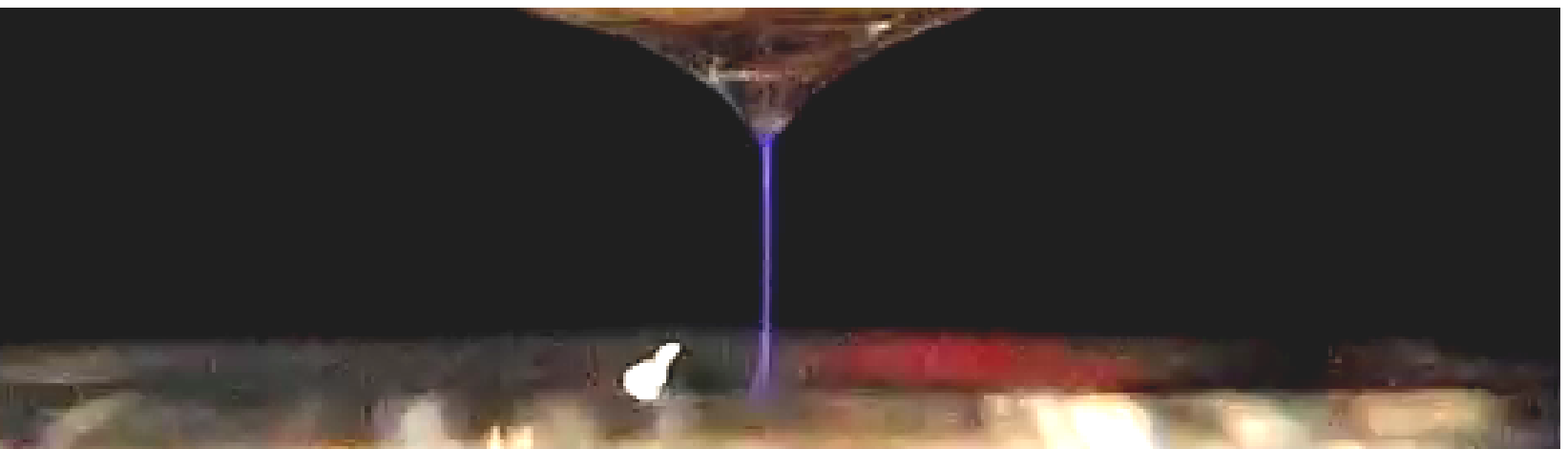}\\
 \includegraphics[width=40mm]{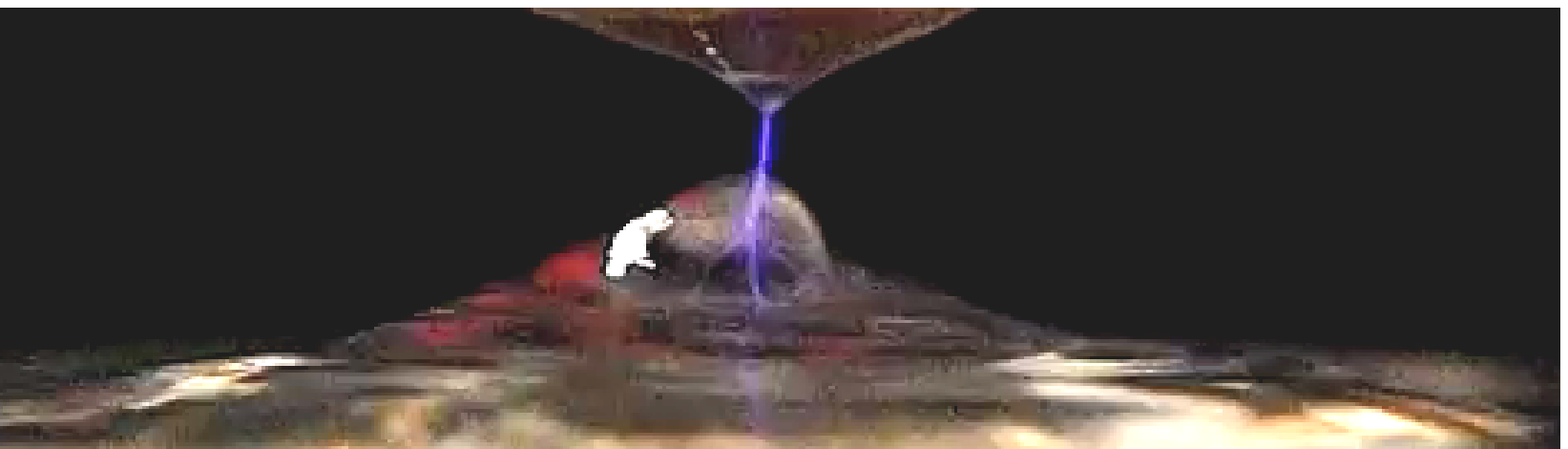}\\
 \includegraphics[width=40mm]{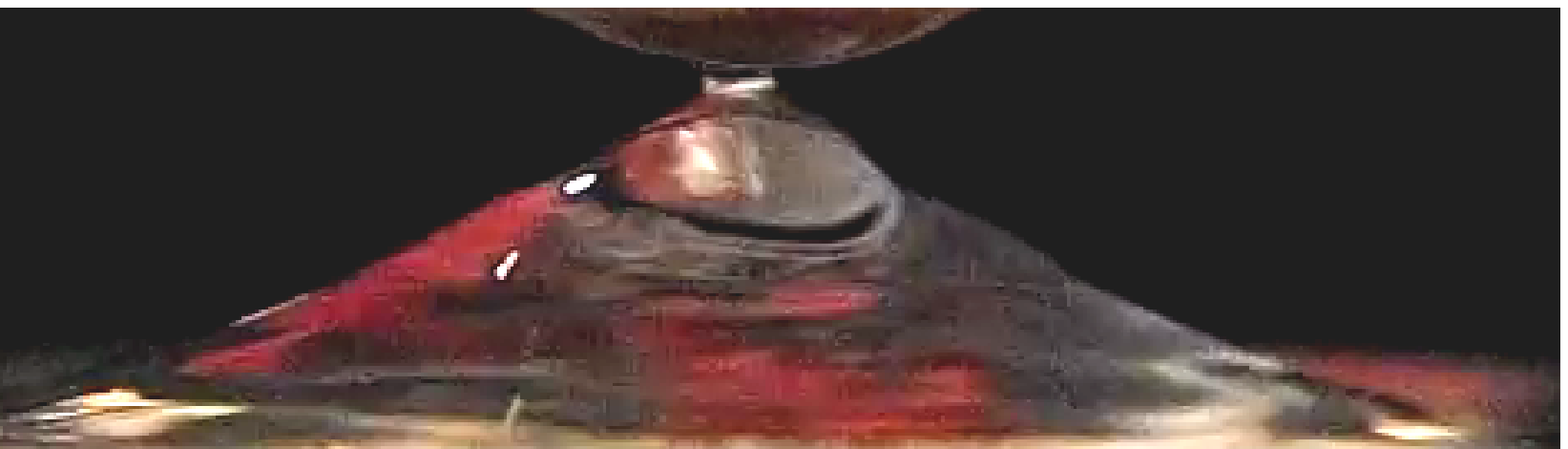}\\
 \includegraphics[width=40mm]{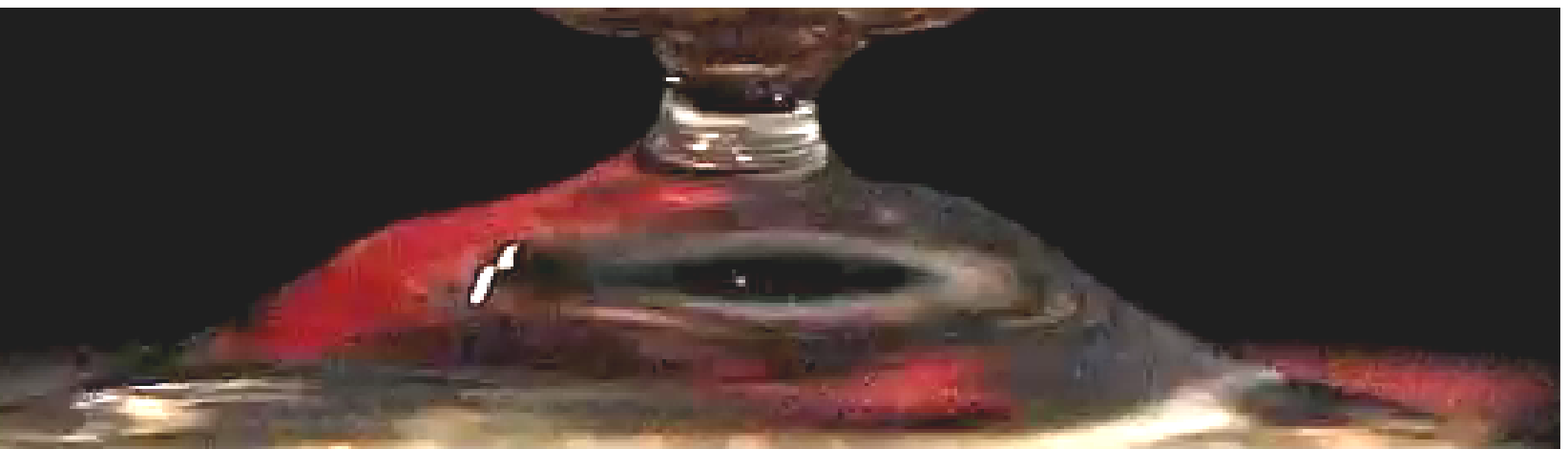}\\
 \includegraphics[width=40mm]{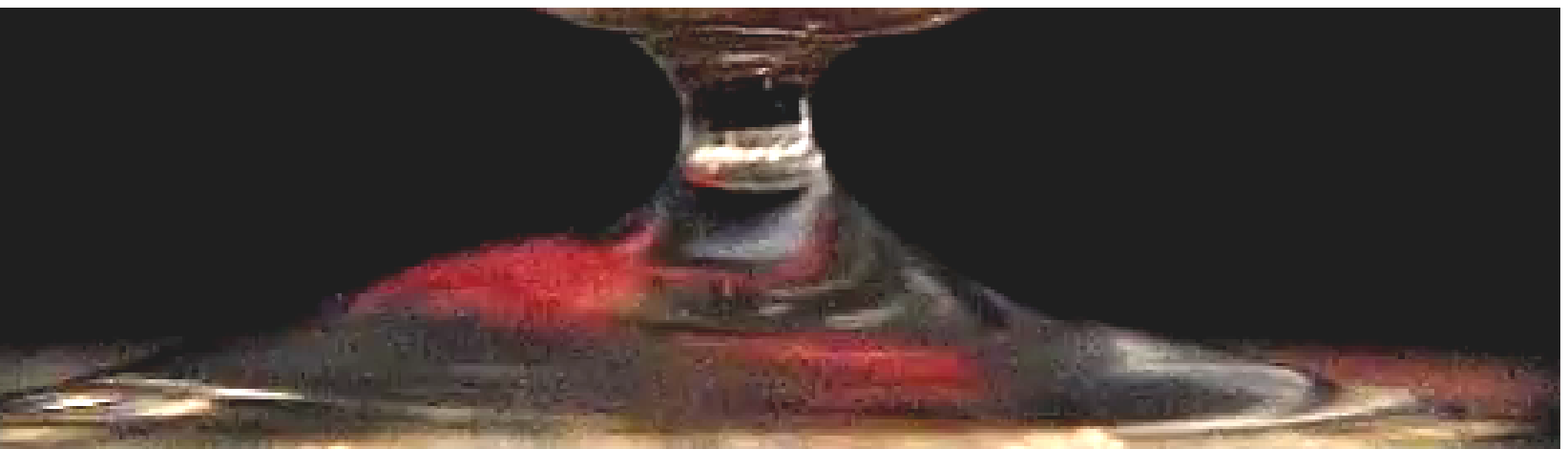}\\
 \includegraphics[width=40mm]{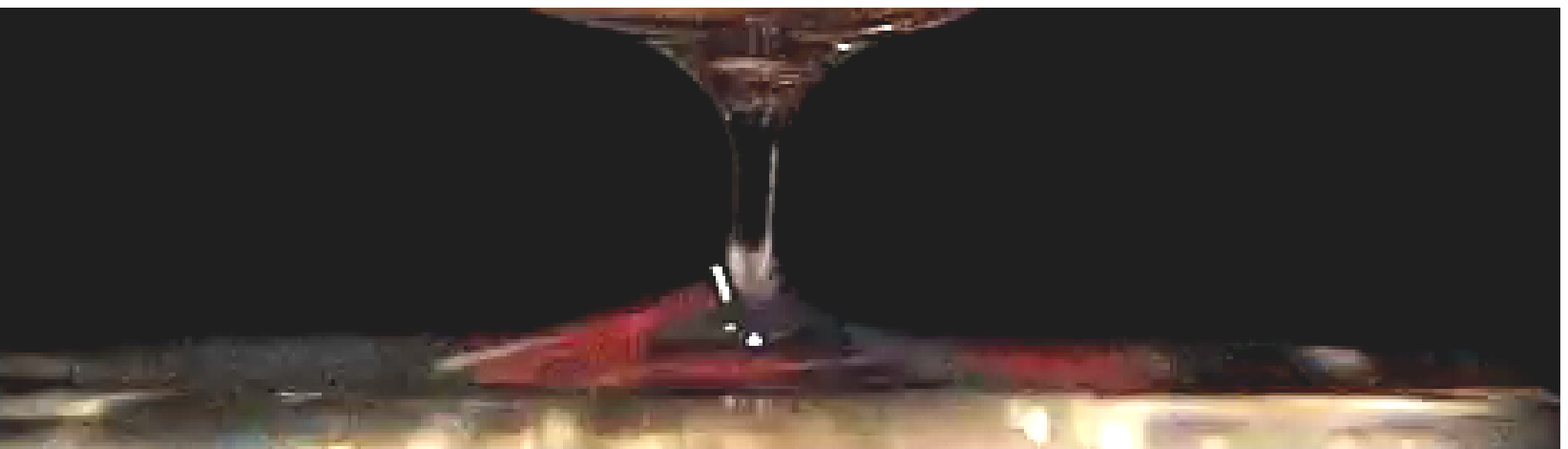}\\
 \includegraphics[width=40mm]{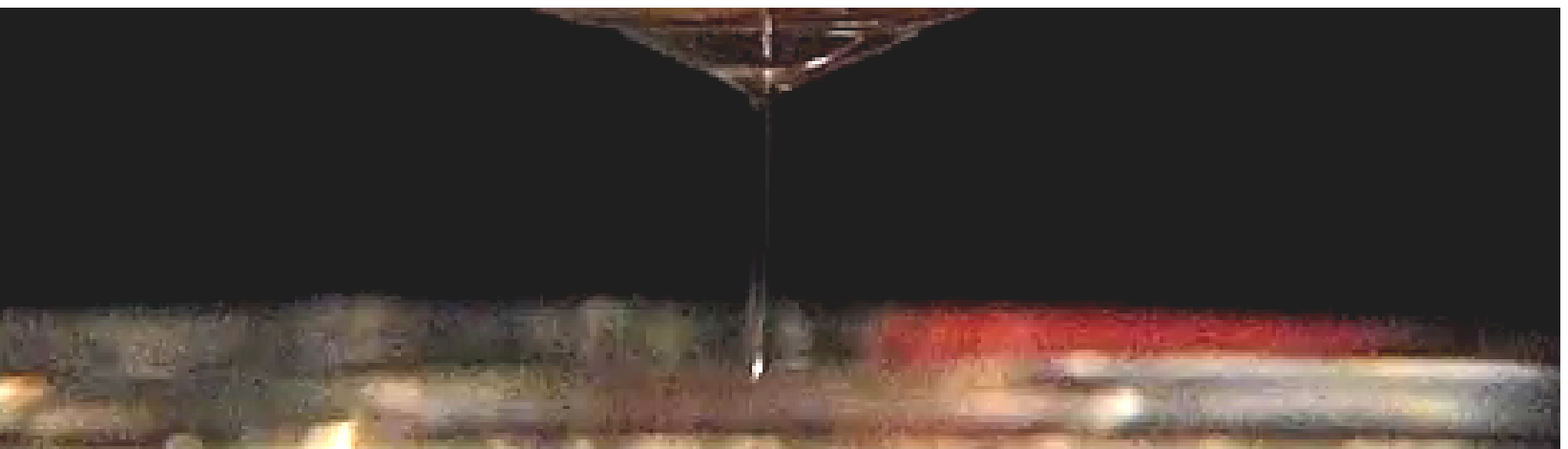}\\
 \includegraphics[width=40mm]{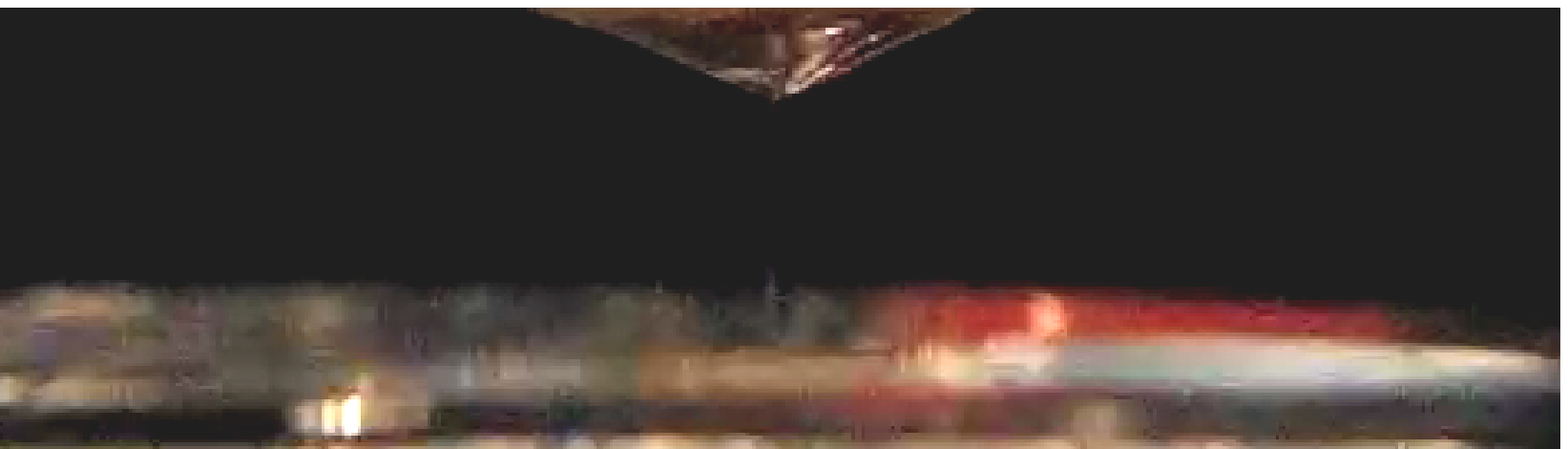}
 \caption{(Color online) Oscillatory motion between destruction and re-establishment of the bridge in presence of a high circuit resistance. $R = 25 M\Omega$, $h = 8mm$, $V_{HV} = 18kV$. Time interval between each frame: $20 \mu s$ and a the motion is repeated. (Video 4)}\label{fig:osc}
\end{figure}

In longer distances between the electrode and water surface and large enough electric fields, the motion changes to a second kind of oscillatory motion (b): destruction and re-establishment of the bridge (Fig. \ref{fig:osc}, Video 4). This happens when the voltage $V_{HV}$ is high enough to raise the water and construct the bridge, but right after its formation electric current increases causing a significant voltage drop and thus the conditions for a stable bridge is no longer valid. So the bridge again falls as there are not enough forces to hold it up. Again with the bridge destruction and reduction of the current, voltage increases and surface charges raise another bridge. The connection phase is along with electric discharges since the field increases as the tip of the electrode approaches the upcoming column of water.

A third kind of motion happens in longer distances (c): when the electric discharges conduct enough current to decrease the field and avoid the formation of the bridge, thus in this case a column of water rises close to the electrode, electrical discharges take place and the column goes back down. Again when the distance gets too far for a discharge to maintain, voltage and electric field increase,leading to another rising column (Video 5). Note that gravitational waves on the free surface of water are effective in the oscillatory motions, as it was observed that the frequency exhibited was also a function of the dimensions of the beaker. The oscillations sometimes needed a starting perturbation, otherwise a stable discharge would take place.

\begin{figure*}
 \centering
 \includegraphics[trim = 0mm 25mm 0mm 25mm, clip, width=0.1\linewidth]{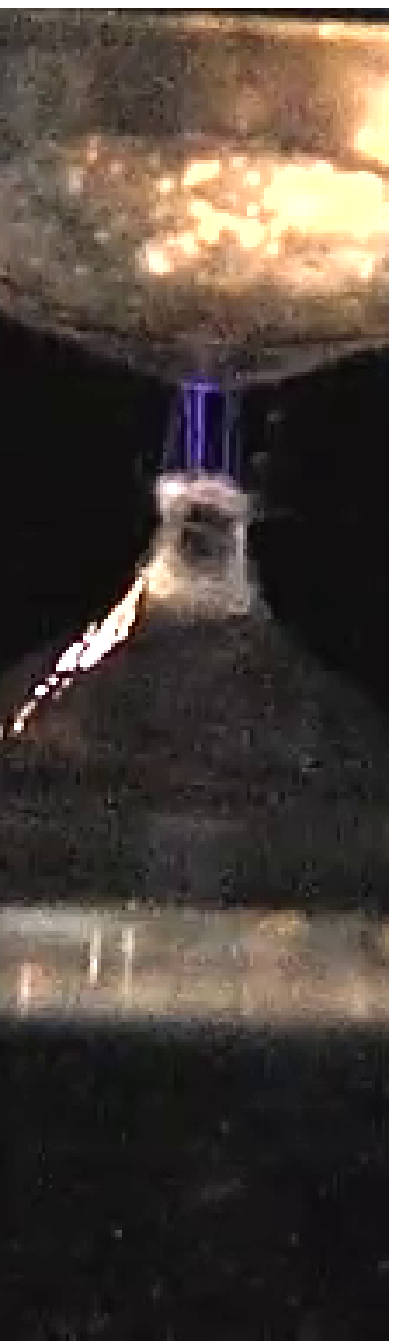}
 \includegraphics[trim = 0mm 25mm 0mm 25mm, clip, width=0.1\linewidth]{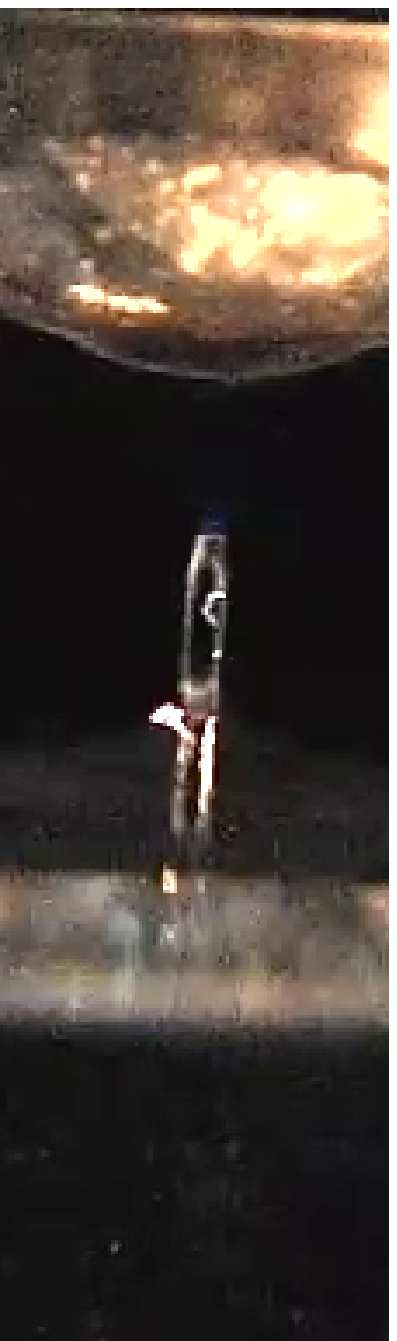}
 \includegraphics[trim = 0mm 25mm 0mm 25mm, clip, width=0.1\linewidth]{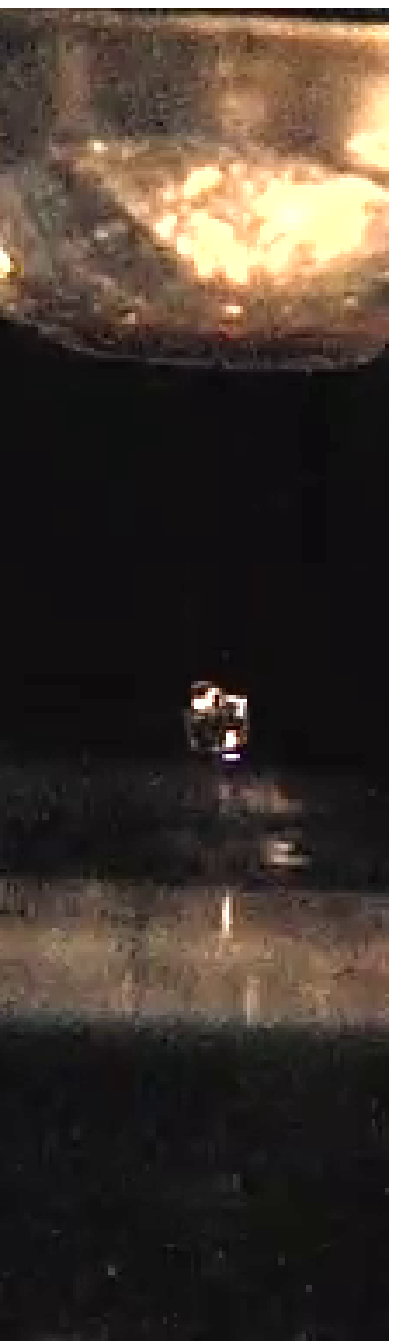}
 \includegraphics[trim = 0mm 25mm 0mm 25mm, clip, width=0.1\linewidth]{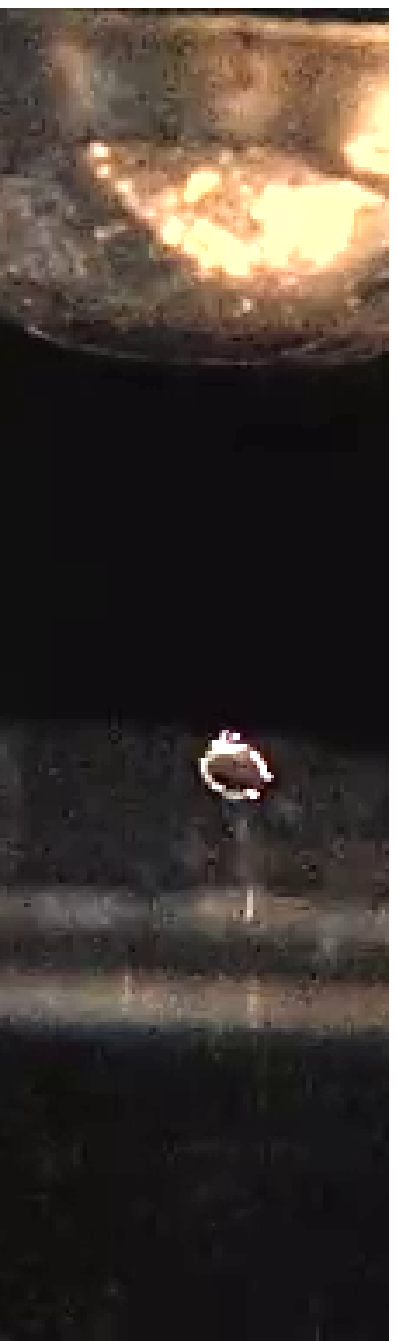}
 \includegraphics[trim = 0mm 25mm 0mm 25mm, clip, width=0.1\linewidth]{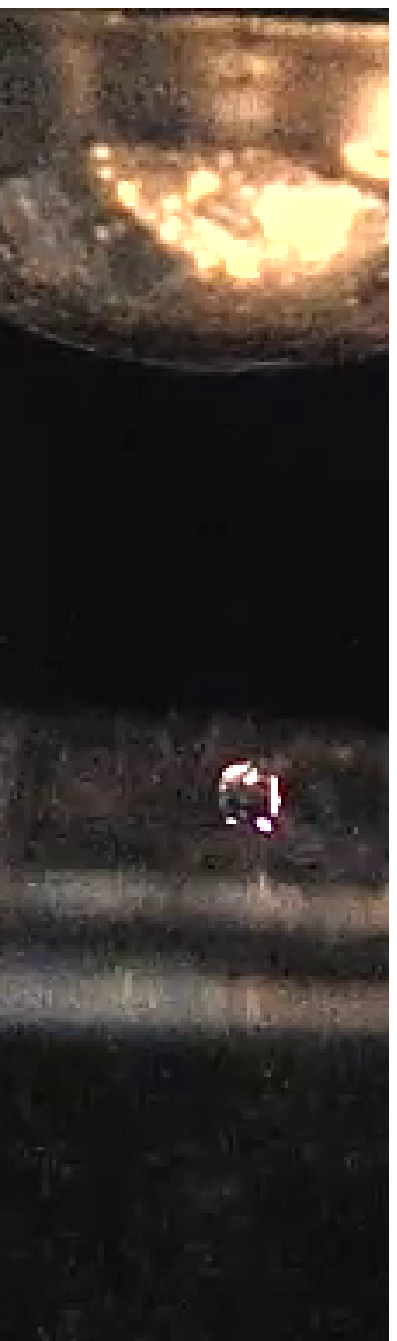}
 \includegraphics[trim = 0mm 25mm 0mm 25mm, clip, width=0.1\linewidth]{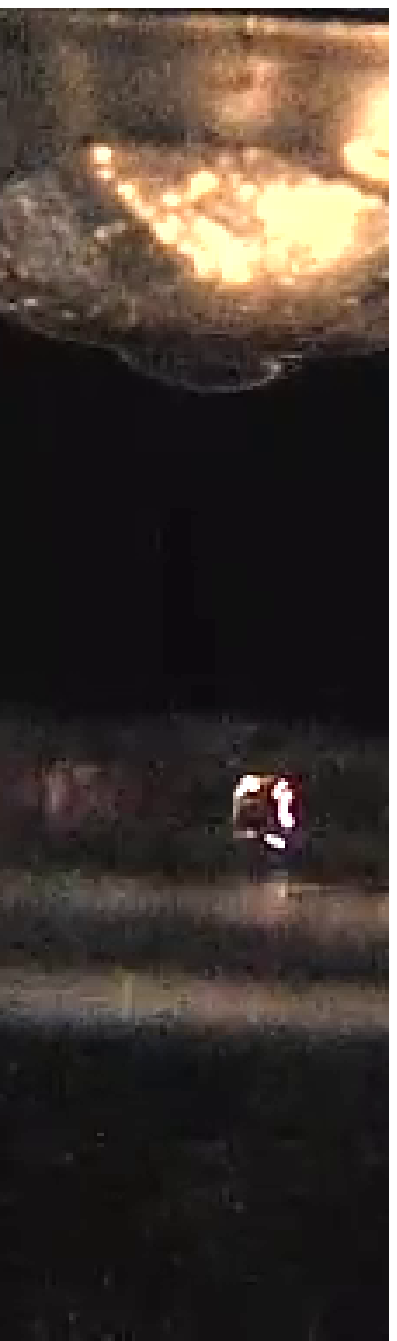}
 \includegraphics[trim = 0mm 25mm 0mm 25mm, clip, width=0.1\linewidth]{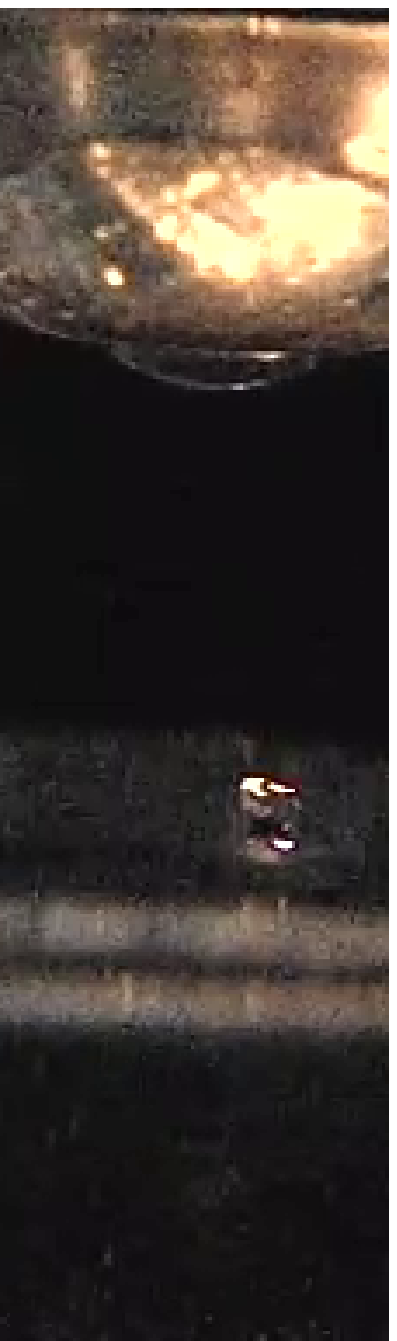}
 \includegraphics[trim = 0mm 25mm 0mm 25mm, clip, width=0.1\linewidth]{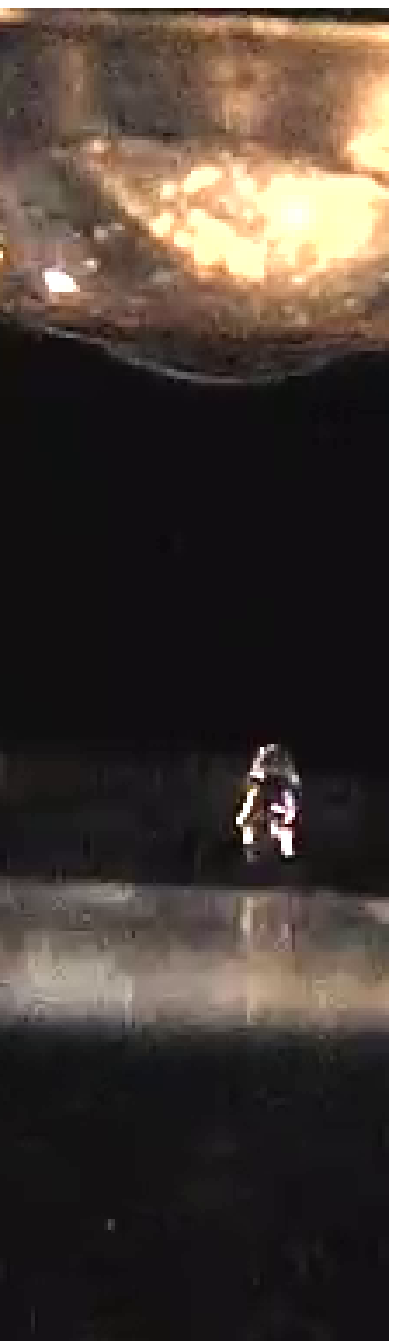}
 \includegraphics[trim = 0mm 25mm 0mm 25mm, clip, width=0.1\linewidth]{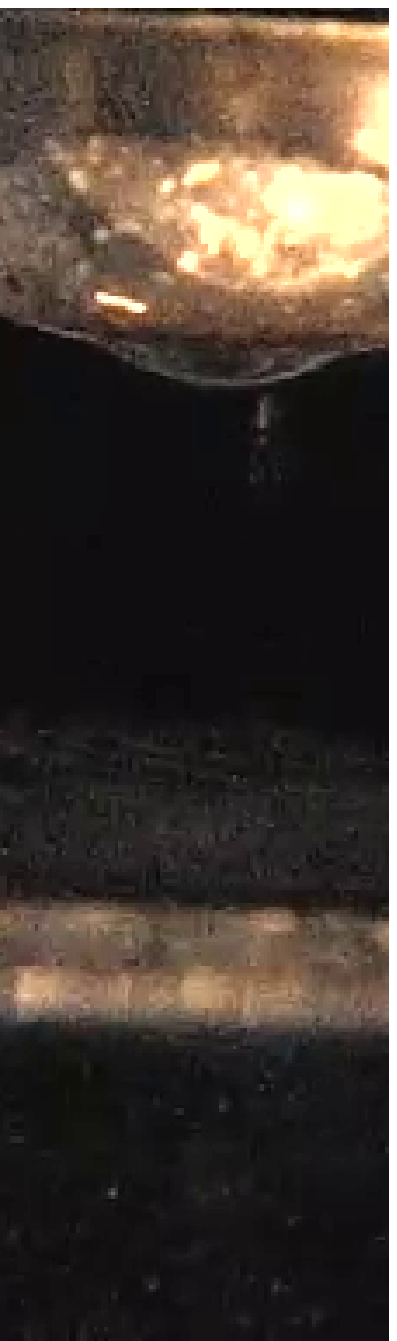} 
 \caption{(Color online) Formation and suspension of a charged macroscopic droplet. The droplet is pulled upwards at the end. $R = 25 M\Omega$, $h = 8mm$, $V_{HV} = 18kV$. Time interval between each frame: $8.3 \mu s$ (Video5)}\label{fig:drop}
\end{figure*}

When the discharges are present in the oscillation, the phenomenon is sometimes along with electrospraying at the tip of the rising column i.e. the formation of charged droplets in the sharp edge of the column. The mechanism is well describes by Taylor \cite{taylor1964disintegration}. We observed that in some cases one or more large macroscopic droplet are formed after a discharge. The droplets are formed after the formation of a slender column which gets separated into droplets as a result of the Plateau-Rayleigh instability. Motion of the droplets shows that a significant upward force is exerted to them. This force is exerted to the droplets by the electric field due to their charge. In different cases the droplets move upward, downward or sometimes almost levitate as the upward electric force magnitude gets close to their weight. Fig. \ref{fig:drop} shows the formation and motion of an almost levitating large droplet, accompanied with a video that shows many cases of drop formation in our experiments.

\section{Conclusion}
Discussions on a horizontal bridge have already led to a diverging range of explanations and this phenomena yet cannot be considered a solved task. Meanwhile the vertical bridge in which geometrical complexities are reduced brings a valuable resource of research for understanding the underlying physics in similar phenomenon.
In this framework, we have observed and experimentally investigated the vertical water bridge. We have observed instabilities in the bridge that lead to time-varying bridge shapes in three different regimes, and these instabilities were observed for the first time. Also the formation of macroscopic droplets which experience an upward force have been observed, proving the charged nature of the droplets. An explanation to these instabilities and features brings a new understanding to the field of electrohydrodynamics.

\section*{Acknowledgments}
We appreciate technical support of Prof. A. Amjadi and thank N. Jafari for his help in experiments. All the experiments were performed in Medical Physics and Laser Lab at Physics department, Sharif University of Technology.

------------------------------------------------

\bibliography{VWB1_EPL}
\bibliographystyle{unsrt}
\end{document}